\documentclass{elsart}  

\usepackage{graphicx}
\usepackage{amsmath}

\begin{document}
%%%%%%%%%%%%%%%%%%%%%%%%%%%%%%%%%%%%%%%%%%%%%%%%%%%%%%%%%%%%%%%%%%%%%%%%%%%%%%%%
\begin{frontmatter}
\title{Uncertainty of polarized gluon distribution from prompt photon production}

\author{M. Hirai \corauthref{cor}}
\corauth[cor]{Present affiliation: 
          Institute of Particle and Nuclear Studies, 
          High Energy Accelerator Research Organization, 
          1-1, Ooho, Tsukuba, Ibaraki, 305-0801, Japan}
\ead{E-mail:mhirai@rarfaxp.riken.jp}

\address{Radiation Laboratory, 
RIKEN (The Institute of Physical and Chemical Research) \\
Wako, Saitama 351-0198, JAPAN }

\begin{abstract}
Constraint of prompt photon data on the polarized gluon distribution
is discussed in terms of uncertainty estimation for 
polarized parton distribution functions (PDFs).
By comparing uncertainty of the double spin asymmetry $A_{LL}^\gamma$
with expected statistical errors at RHIC, 
we found that the gluon distribution is effectively constrained 
in the region $0.04<x_{_T}<0.2$ 
with the data at transverse momentum $p_{_T}=10-20$ GeV 
for center-of-mass energies $\sqrt{s}=200$ and 500 GeV.
\end{abstract}

\end{frontmatter}
%%%%%%%%%%%%%%%%%%%%%%%%%%%%%%%%%%%%%%%%%%%%%%%%%%%%%%%%%%%%%%%%%%%%%%%%%%%%%%%%
%%% Introduction %%%%
\section{Introduction}
By the recent global analyses with the experimental data 
for polarized deep inelastic scattering (DIS),
the polarized quark and antiquark distributions are determined well 
\cite{AAC03,BB,LSS,GRSV}.
These distributions are obtained with enough accuracy to indicate that 
the quark spin content is smaller than prediction of naive quark model; 
$\Delta \Sigma=0.1\sim0.3 \ (\ne1)$.
These polarized parton distribution functions (PDFs) 
reproduce well the experimental data;
however, the polarized gluon distribution $\Delta g(x)$ cannot be constrained 
because of indirect and small contribution through $Q^2$ evolution 
and higher order correction at next-to-leading order (NLO). 
Furthermore, PDF uncertainty estimation indicated 
large uncertainty of the gluon distribution. 
It implies difficulty of the $\Delta g(x)$ determination 
with only the polarized DIS data. 

As a probe for the polarized gluon distribution, 
prompt photons will be measured 
by the longitudinally polarized proton-proton collider at RHIC \cite{RHIC-S}.
The gluon distribution contributes directly 
in the quark-gluon compton process $(qg\to\gamma q)$ at leading order (LO), 
and the process dominates in the whole $p_{_T}$ region.
The future asymmetry data contain useful information 
for clarifying the gluon contribution to the nucleon spin. 
Therefore, we are interested in the influence of the prompt photon data on 
the $\Delta g(x)$ determination by the polarized PDF analysis.

%%%% motivation of this paper
In this paper, 
we consider constraint of prompt photon data
on the polarized gluon distribution.
For evaluating the data constraint, 
we will compare the uncertainty of the spin asymmetry
with the expected statistical error by the RHIC experiments.
The asymmetry uncertainty coming from the polarized PDFs 
is estimated by the Hessian method, 
and it is comparable with the measurement error. 
In this comparison, 
the statistical error of the spin asymmetry plays a role of 
constriction for the $\Delta g$ uncertainty via the asymmetry.
In practice, it is indicated that 
uncertainties of the polarized PDFs can be reduced 
by including new precise data for the polarized DIS 
in the Asymmetry Analysis Collaboration (AAC) \cite{AAC03}. 
Therefore, the same thing is expected 
by including the future data for prompt photon production. 

%%%%%%%%%%%%%%%%%%%%%%%%%%%%%%%%%%%%%%%%%%%%%%%%%%%%%%%%%%%%%%%%%%%%%%%%%%%%%%%
%%% uncertainty of spin asymmetry and expected statistical error %%%
\section{Uncertainty of the spin asymmetry}
The spin asymmetry $A_{LL}^{\gamma}$ is defined 
as a ratio of the polarized and unpolarized cross sections:
$A_{LL}^\gamma=\Delta \sigma^\gamma/\sigma^\gamma$.
By the factorization theorem, 
the polarized cross section $\Delta \sigma^{\gamma} (\vec{p}_A \vec{p}_B\to\gamma X)$ 
as a function of the transverse momentum $p_{_T}$ is expressed by
\begin{align}
    \frac{d\Delta \sigma^\gamma}{d p_{_T}} &= 
        \sum_{a,b} \int_{\eta-bin} d\eta \int d x_a \int d x_b  \\ \nonumber 
     & \times \Delta f_a^A(x_a,\mu_{_F})\ \Delta f_b^B(x_b,\mu_{_F}) \\ \nonumber
     & \times \frac{d \Delta \hat{\sigma}_{ab}^\gamma}
             {d p_{_T} d\eta} (x_a,x_b,\sqrt{s},p_{_T},\eta,\mu_{_R},\mu_{_F}) \ ,
    \label{eq:xsec}
\end{align}
where $\Delta f_a(x)$ is the polarized PDF of the parton $a$. 
We choose the AAC03 PDF set \cite{AAC03}.
$\Delta \hat{\sigma}_{ab}^\gamma$ is the partonic cross section 
($a+b\to \gamma + X$). 
In order to reduce theoretical uncertainty 
from the scale dependence of the cross section, 
the NLO corrections are taken into account. 
The NLO partonic cross sections for prompt photon production 
are completely known \cite{PPPNLO}.
The renormalization and factorization scales are chosen 
the scale $\mu_{_F}=\mu_{_R}=p_{_T}$.
In addition, 
the cross section is integrated over the rapidity bin $|\eta|<0.35$, 
which corresponds to the acceptance of the PHENIX detector.
The unpolarized cross section $\sigma^\gamma$ 
is similarly computed with unpolarized PDFs and 
partonic cross sections \cite{PPPNLO}. 
We choose the GRV98 unpolarized PDF set \cite{GRV98}, 
which is also used in the AAC03 analysis.
These cross sections are numerically calculated 
at center-of-mass (c.m.) energy $\sqrt{s}=200$ and 500 GeV, respectively.

In this study, the contribution from fragmentation is neglected.
The contribution is associated with the collinear process for 
a scattered parton into a photon, 
and it can be diminished by using an isolation cut 
on the measured photon \cite{Frixione}.
And so we should consider the isolation cut 
in this analysis \cite{ISCut}. 
For examining an effect on the polarized PDF uncertainties, 
we calculate the cross section for 
inclusive direct photon production process.

% The uncertainty of the polarized cross section
The asymmetry uncertainty is obtained from uncertainty 
of the polarized cross section: 
$\delta A_{LL}^\gamma=\delta \Delta \sigma^\gamma/\sigma^\gamma$.
The uncertainty $\delta \Delta \sigma^\gamma$ comes from 
the polarized PDF uncertainties, 
and can be estimated by the Hessian method.
The method based on eigenvectors of the diagonalized Hessian matrix 
is developed by CTEQ collaboration \cite{CTEQ-H}, 
and it is applied to estimate uncertainties of unpolarized PDFs \cite{unPDF-E}. 
We used the basic method, which is generally used 
by the polarized PDF analyses \cite{AAC03,BB,LSS}. 
The uncertainty is given by 
\begin{equation}
    [\delta \Delta \sigma^\gamma]^2=
       \Delta \chi^2 \sum_{i,j}
       \left( \frac{\partial \Delta \sigma^\gamma(p_{_T})}
                   {\partial a_i} \right)
       H_{ij}^{-1}
       \left( \frac{\partial \Delta \sigma^\gamma(p_{_T})}
                   {\partial a_j} \right) \ ,
    \label{eq:erroe-M}
\end{equation}
where $a_i$ are optimized parameters in the polarized PDFs.
$H_{ij}$ is the Hessian matrix
which has information on the parameter errors 
and correlation between each parameter. 
The gradient terms of the cross section 
$\partial \Delta \sigma^\gamma(p_{_T})/\partial a_i$ 
are obtained as follows:
\begin{align} 
\label{eq:gradf}
    \frac{\partial \Delta \sigma^\gamma}{\partial a_i} &=
         \sum_{a,b} \int_{bin} d\eta \int d x_a \int d x_a \\ \nonumber
      &  \times \left[ 
               \frac{\partial \Delta f_a^A(x_a)}{\partial a_i} \Delta f_b^B(x_b)
              +\Delta f_a^A(x_a)\frac{\partial \Delta f_b^B(x_b)}{\partial a_i} 
               \right] \\ \nonumber
      & \times \frac{d \Delta \hat{\sigma}_{ab}^\gamma}{d p_{_T} d\eta} 
                     (x_a,x_b,\sqrt{s},p_{_T},\eta,\mu_{_R},\mu_{_F}) \ .
\end{align}
The gradient terms of the polarized PDFs can be derived analytically 
at initial scale, and these terms are numerically evolved 
to arbitrary scale $Q^2(=p_{_T}^2)$ by the DGLAP equation.
Furthermore, 
the value of $\Delta \chi^2$ determines a confidence level of the uncertainty. 
It is obtained so that the level corresponds to 1$\sigma$ error 
of the normal distribution \cite{AAC03}. 
The uncertainty therefore can be directly compared with 
the statistical error of experimental data.

%%% expected statistical error at RHIC
For comparison with the asymmetry uncertainty, 
the statistical error of the spin asymmetry is estimated by
\begin{equation}
\label{eq:staerr}
    \delta A_{LL}^{exp}=\frac{1}{P^2\sqrt{\mathcal{L}_{int} \sigma}} \ ,
\end{equation}
where $P$ is the beam polarization, 
$\mathcal{L}_{int}$ is the integrated luminosity, 
and $\sigma$ is the unpolarized cross section integrated over the $p_{_T}$ bin. 
In this study, the $\sigma$ is computed by the bin size of 5 GeV interval, 
and it is taken the same bin size for both c.m. energies 
$\sqrt{s}=200$ and $500$ GeV.
Furthermore, other values are chosed design values at RHIC \cite{RHIC-S}: 
$P=0.7$ and $\mathcal{L}_{int}=320 \ (800)$ pb$^{-1}$ 
for $\sqrt{s}= 200\ (500)$ GeV. 

%%%%%%%%%%%%%%%%%%%%%%%%%%%%%%%%%%%%%%%%%%%%%%%%%%%%%%%%%%%%%%%%%%%%%%%%%%%%%%%
\section{Constraint of prompt photon data on $\Delta g(x)$}
%%% the spin asymmetry and its uncertainty
%%%%%%%%%%%%%%%%%%%%%%%%%%%%%%%% figure %%%%%%%%%%%%%%%%%%%%%%%%%%%%%%%%%%%%%%%%
\begin{figure}[b]
\begin{center}
        \includegraphics*[scale=0.42]{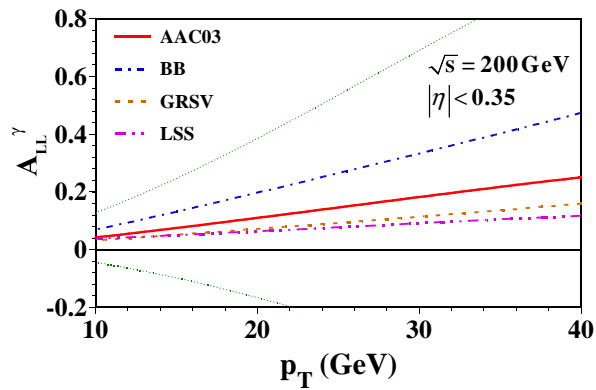} 
\caption{\label{fig:asymmetry}
Comparison of the predicted spin asymmetries by different polarized PDF sets; 
AAC03(NLO), BB (ISET=3), GRSV01 (standard scenario), 
and LSS ($\overline{\rm MS}$ scheme).
The dotted curves show the asymmetry uncertainty 
from the PDF uncertainties of the AAC03 set. 
}
\end{center}
\end{figure}
%%%%%%%%%%%%%%%%%%%%%%%%%%%%%%%% figure %%%%%%%%%%%%%%%%%%%%%%%%%%%%%%%%%%%%%%%%

First, we discuss predicted spin asymmetry and its uncertainty 
at $\sqrt{s}=200$ GeV.
In Fig. \ref{fig:asymmetry}, 
the spin asymmetry by the AAC03 PDF set is compared to 
those by polarized PDF sets of 
BB (ISET=3) \cite{BB}, 
GRSV01 (standard scenario)\cite{GRSV}, 
and LSS ($\overline{\rm MS}$ scheme) \cite{LSS}. 
These analyses used almost the same experimental data sets for the polarized DIS, 
and they obtained good agreements with the data.
However, there are significant differences of the gluon distributions among them.
\footnote{See, for example, Ref. \cite{AAC03,BB}.}
These differences are obviously reflected in variations of the predicted asymmetries.
The prompt photon process is sensitive to the behavior of the gluon distribution. 
Moreover, the asymmetry uncertainty is indicated in the same figure. 
Dotted curves show the uncertainty which comes from the polarized PDF uncertainties 
obtained by the AAC analysis with the polarized DIS data.
We find that these predicted asymmetries are within the large uncertainty.
These variations are caused by weak constraint of the polarized DIS data 
on the gluon distribution.
The prompt photon data therefore are required for 
reducing this asymmetry uncertainty.

\begin{figure}[b]
\begin{center}
        \includegraphics*[scale=0.42]{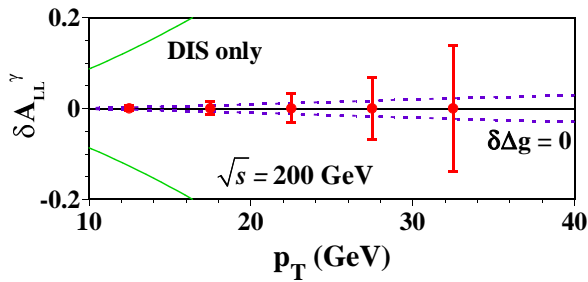} 
\caption{\label{fig:dgoff}
Comparison of the asymmetry uncertainty $\delta A_{LL}^\gamma$
with the expected statistical errors for $\sqrt{s}=200$ GeV.
The dashed curves show the asymmetry uncertainty except the gluon uncertainty: 
$\delta \Delta g(x)=0$.
}
\end{center}
\end{figure}

%%% Comparison asymmetry uncertainty with expected statistical error %%%
% Component of the asymmetry uncertainty
In order to evaluate the gluon contribution to the asymmetry uncertainty, 
we compute the asymmetry uncertainty 
excluding the $\Delta g(x)$ uncertainty 
by assuming $\partial \Delta g(x)/\partial a_g=0$ in eq. (\ref{eq:gradf}). 
Figure \ref{fig:dgoff} shows the asymmetry uncertainties for $\sqrt{s}= 200$ GeV.
The solid curves show the current uncertainty by the AAC analysis, 
and the dotted curves do the asymmetry uncertainty 
except the $\Delta g(x)$ uncertainty. 
The significant reduction of the uncertainty indicates that 
the $\Delta g(x)$ uncertainty is the dominant contribution to the current uncertainty.

In addition, 
the asymmetry uncertainty is compared to the expected statistical errors at RHIC. 
The current uncertainty is much larger than these statistical errors. 
If these data are included in the global analysis, 
the uncertainty is roughly reduced to these errors. 
This suggests that the $\Delta g(x)$ uncertainty is mainly improved. 
The prompt photon data therefore have strong constraint on the gluon distribution. 

%%%%%%%%%%%%%%%%%%%%%%%%%%%%%%%% figure %%%%%%%%%%%%%%%%%%%%%%%%%%%%%%%%%%%%%%%%
%%%%%%%%%%%%%%%%%%%%%%%%%%%%%%%% figure %%%%%%%%%%%%%%%%%%%%%%%%%%%%%%%%%%%%%%%%

% The data constraint on the quark and antiquark uncertainties
On the other hand, 
the data constraint on the quark and antiquark distributions is very weak.
This is because that
the asymmetry uncertainty without the $\Delta g(x)$ uncertainty, 
which is composed of the $\Delta q(x)$ and $\Delta \bar{q}(x)$ uncertainties, 
is significantly less than the statistical errors. 
Therefore, the data with such errors do not directly affect 
improvements of these uncertainties.

However, as far as the antiquark is concerned, 
the uncertainty can be indirectly reduced 
because the antiquark distribution is strongly correlated with 
the gluon distribution in the global analysis.
In practice, 
reduction of the antiquark uncertainty via the error correlation is indicated 
by the analysis with the fixed $\Delta g(x)=0$ at initial scale \cite{AAC03}. 
This fact suggests that the constraint on the gluon distribution 
indirectly affects the $\Delta \bar{q}(x)$ determination. 
In particular, it is not neglected 
when we perform flavor decomposition of the antiquark distributions.

%%% Constricted gluon uncertainty from the statistical error %%%
Next, we estimate a constraint factor for the $\Delta g(x)$ uncertainty. 
By multiplying the gradient terms for the gluon $\partial \Delta g(x)/\partial a_g$ 
by the factor in eq. (\ref{eq:gradf}), 
the constricted uncertainty of the asymmetry is defined. 
From comparison of the asymmetry uncertainty with 
the expected statistical errors for $\sqrt{s}=200$ in the region $10<p_{_T}<20$ GeV, 
the obtained factor is $1/18$. 

In this study, 
the factors for the $\Delta q(x)$ and $\Delta \bar{q}(x)$ uncertainties 
are neglected. 
This is simply because that 
these uncertainties are not directly constricted by these data.
Moreover, the correlation effect on the $\Delta \bar{q}(x)$ uncertainty 
is not taken into account. 
The effect cannot be evaluated without including experimental data 
in the global analysis. 
Since the $\Delta \bar{q}(x)$ contribution to the asymmetry uncertainty 
is already small, the uncertainty will be slightly modified in this process.

\begin{figure}[b!]
\begin{center}
        \includegraphics*[scale=0.42]{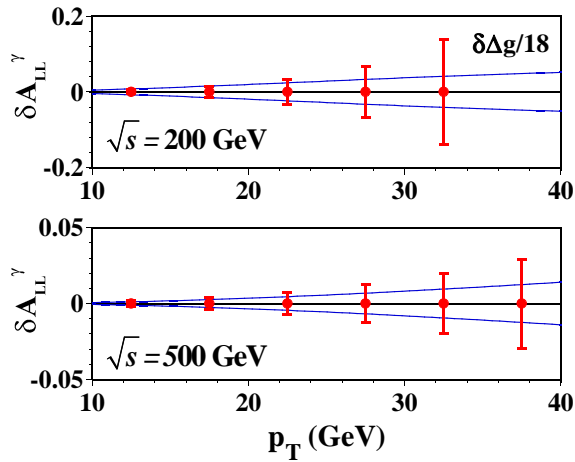} 
\caption{\label{fig:constrict}
Comparison of the constricted asymmetry uncertainties with 
the expected statistical errors for $\sqrt{s}=200$ and $500$ GeV, respectively.
The solid curves show the constricted asymmetry uncertainty 
with the factor $1/18$ for the $\Delta g(x)$ uncertainty.
}
\end{center}
\end{figure}

% the data constraint for $\sqrt{s}=200$ GeV
In Fig. \ref{fig:constrict}, 
the constricted asymmetry uncertainties are compared with
the expected statistical errors for $\sqrt{s}=200$ and $500$ GeV, respectively.
The solid curves show the asymmetry uncertainties 
which are obtained from 
the $\Delta g(x)$ uncertainty multiplied by the constraint factor, 
and involve the $\Delta q(x)$ and $\Delta \bar{q}(x)$ uncertainties. 
If $x_{_T}(=2p_{_T}/\sqrt{s})$ can be approximated by the Bjorken $x(=x_a=x_b)$ 
around central rapidity region, 
the data for $\sqrt{s}=200$ GeV in the region $10<p_{_T}<20$ GeV 
constrain the gluon distribution in the range $0.1<x<0.2$. 
Although this comparison is in ideal condition 
that these data are put on the predicted asymmetry, 
this fact agrees with the results of 
the trial analysis including pseudo-data for $A_{LL}^\gamma$ in Ref. \cite{SV}. 
These data are useful in determining the gluon distribution.

In the region $p_{_T}>20$ GeV, these data have rather weak constraint 
since the unpolarized cross section rapidly decreases as $p_{_T}$ increases.
The statistical errors depend on the $p_{_T}$ bin size 
computing $\sigma$ in eq. (\ref{eq:staerr}).
We should be careful about taking the bin size for the high $p_{_T}$ data 
in order to constrain equally the gluon distribution over a wide $x$ region.
%%%%%%%%%%%%%%%%%%%%%%%%%%%%%%%% figure %%%%%%%%%%%%%%%%%%%%%%%%%%%%%%%%%%%%%%%%
%\begin{figure}[t]
%%%%%%%%%%%%%%%%%%%%%%%%%%%%%%%% figure %%%%%%%%%%%%%%%%%%%%%%%%%%%%%%%%%%%%%%%%

% the data constraint for $\sqrt{s}=500$ GeV
For the comparison at $\sqrt{s}=500$ GeV, we find similar behavior.
In the region $10<p_{_T}<20$ GeV,
the asymmetry uncertainty roughly corresponds to the statistical errors. 
This indicates that these data have the same constraint as 
those for $\sqrt{s}=200$ GeV, 
and constrain the gluon distribution in the range $0.04<x<0.08$.
Above the region, 
the statistical errors are larger than the asymmetry uncertainty. 
It is noteworthy to mention here that 
the data constraint for $\sqrt{s}=500$ GeV is weaker than 
that for $\sqrt{s}=200$ GeV  in the region $10<p_{_T}<20$ GeV 
in spite of covering the same $x_{_T}$ region.

%%%%%%%%%%%%%%%%%%%%%%%%%%%%%%%% figure %%%%%%%%%%%%%%%%%%%%%%%%%%%%%%%%%%%%%%%%
\begin{figure}[b]
\begin{center}
        \includegraphics*[scale=0.40]{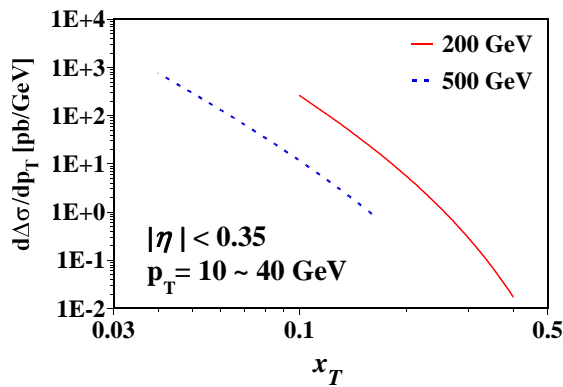} 
\caption{\label{fig:xsec_xt}
The unpolarized cross sections for $\sqrt{s}=200$ and 500 GeV.
}
\end{center}
\end{figure}
%%%%%%%%%%%%%%%%%%%%%%%%%%%%%%%% figure %%%%%%%%%%%%%%%%%%%%%%%%%%%%%%%%%%%%%%%%
The reason for the weak constraint is that 
the unpolarized cross section for $\sqrt{s}=500$ GeV is less than 
that for $\sqrt{s}=200$ GeV in the same $x_{_T}$ region, 
and the integrated luminosity is still insufficient 
to provide the enough constraint at high $p_{_T}$. 
Figure \ref{fig:xsec_xt} shows the comparison of 
unpolarized cross sections for both c.m. energies.
These cross sections are indicated as a function of $x_{_T}$.
In the region $x_{_T}>0.1$, 
the cross section for $\sqrt{s}=500$ GeV is below one for $200$ GeV, 
and indicates similar behavior. 
In order to obtain equal constraint at the same $x_{_T}$, 
we need more luminosity than the design value at $\sqrt{s}=500$ GeV. 
Therefore, the experimental data for $\sqrt{s}=500$ GeV are required 
as constraint on the gluon distribution at low $x$. 
The medium-$x$ behavior should be determined by using the data for $200$ GeV.

%%% Constraint gluon uncertainty from prompt photon production at RHIC
%%%%%%%%%%%%%%%%%%%%%%%%%%%%%%%% figure %%%%%%%%%%%%%%%%%%%%%%%%%%%%%%%%%%%%%%%%
\begin{figure}[t]
\begin{center}
        \includegraphics*[scale=0.42]{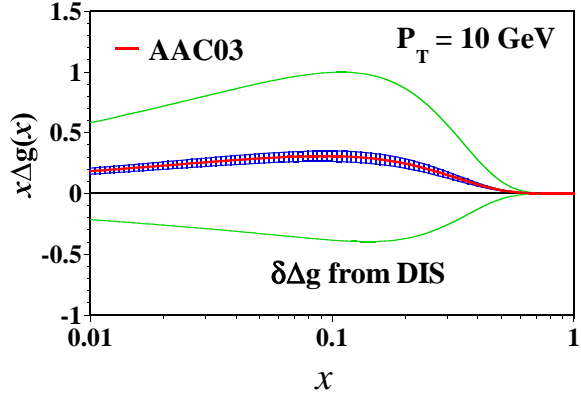} 
\caption{\label{fig:xdg}
The polarized gluon distribution and its uncertainty at $p_{_T}=10$ GeV.
Shaded area is the constricted gluon uncertainty.
}
\end{center}
\end{figure}
%%%%%%%%%%%%%%%%%%%%%%%%%%%%%%%% figure %%%%%%%%%%%%%%%%%%%%%%%%%%%%%%%%%%%%%%%%
Finally, let us turn to the $\Delta g(x)$ uncertainty 
from the prompt photon data at RHIC.
Figure \ref{fig:xdg} shows 
the polarized gluon distribution and its uncertainties at $p_{_T}=10$ GeV.
The solid curves show the $\Delta g(x)$ uncertainty from the polarized DIS data, 
and the shaded area shows the constricted uncertainty 
by comparison with the expected statistical errors. 
The uncertainty estimated by the constraint factor is reliable 
in the range $0.04<x<0.2$. As a practical estimation, 
the uncertainty broadens gradually in the low- and medium-$x$ regions 
where data do not exist.
\footnote{See, for example, Fig. 5b in Ref. \cite{SV}.}
Moreover, we note that 
increasing the asymmetry uncertainty with $p_{_T}$ is due to 
uncertainty of the ratio of the polarized and unpolarized 
gluon distributions: $\delta \Delta g(x)/g(x)$. 
The uncertainty significantly increases with $x$ 
due to lack of precise data for polarized DIS at large $x$.
The $\Delta g(x)$ uncertainty becomes the same order of magnitude
as the quark uncertainty from the DIS data.
By including future asymmetry data for the prompt photon process, 
the gluon distribution can be obtained with sufficient accuracy.

% Comment of the gluon polarization
In this study, polarization of the gluon distribution is not discussed. 
The current uncertainty in Fig. \ref{fig:xdg} indicates that 
we cannot rule out the possibility of zero or negative polarization.
As another probe for the gluon distribution, 
the double spin asymmetry for $\pi^0$ production has recently been reported 
by the PHENIX collaboration \cite{PHENIX_pi0}. 
Since $gg \to \pi^0 X$ subprocess dominates at low $p_{_T}$, 
the cross section depends on $(\Delta g)^2$; 
therefore, the process is not sensitive to the sign of the gluon polarization.
On the other hand, 
the prompt photon production which the qg compton process dominates
is sensitive to the sign in the whole $p_{_T}$ region. 
The gluon polarization is obviously reflected in the spin asymmetry. 
In this sense, the role of prompt photon data is of prime importance 
for determination of the gluon polarization.

%%%%%%%%%%%%%%%%%%%%%%%%%%%%%%%%%%%%%%%%%%%%%%%%%%%%%%%%%%%%%%%%%%%%%%%%%%%%%%%
\section{Summary}
In this paper, 
we have considered the uncertainty of the polarized gluon distribution 
for prompt photon production at RHIC.
The uncertainty of the double  spin asymmetry is estimated 
by the Hessian method.
The asymmetry uncertainty mostly comes from the $\Delta g(x)$ uncertainty.
The large uncertainty implies the weak constraint of the polarized DIS data 
on the gluon distribution. 
By comparison with the expected statistical errors at RHIC, 
we indicate that the prompt photon data have the strong constraint on it.
Furthermore, we suggest that the prompt photon data in the region $10<p_{_T}<20$ GeV 
effectively constrain the gluon distribution. 
The data of both c.m. energies constrain it in the following regions:
$0.04<x<0.08$ at $\sqrt{s}=500$ GeV, and $0.1<x<0.2$ at $200$ GeV.
For clarifying the gluon contribution $\Delta g(\equiv \int_0^1dx \Delta g(x))$, 
the data covering a wide range of $x$ are required.
These experiments therefore play an important role in 
the $\Delta g(x)$ determination.

%%%%%%%%%%%%%%%%%%%%%%%%%%%%%%%%%%%%%%%%%%%%%%%%%%%%%%%%%%%%%%%%%%%%%%%%%%%%%%%
\section*{Acknowledgements}
The author would like to thank N. Saito for helpful discussion 
and useful comments. 

%%%%%%%%%%%%%%%%%%%%%%%%%%%%%% reference %%%%%%%%%%%%%%%%%%%%%%%%%%%%%%%%%%%%%%
\def\PR#1#2#3{Phys. Rev. {#1} (#3) #2}
\def\PRL#1#2#3{Phys. Rev. Lett. {#1} (#3) #2}
\def\PLB#1#2#3{Phys. Lett. B {#1} (#3) #2}
\def\NP#1#2#3{Nucl. Phys. {#1} (#3) #2}
\def\EPJ#1#2#3{Eur. Phys. J {#1} (#3) #2}
%%%%%%%%%%%%%%%%%%%%%%%%%%%%%%%%%%%%%%%%%%%%%%%%%%%%%%%%%%%%%%%%%%%%%

%%%%%%%%%%%%%%%%%%%%%%%%%%%%%%%%%%%%%%%%%%%%%%%%%%%%%%%%%%%%%%%%%%%%%%%%%%%%%%%
\end{document}